%% file: main.tex
\documentclass{ifacconf}
\usepackage{cite}
\usepackage{amsmath,amssymb,amsfonts}
\usepackage{algorithmic}
\usepackage{graphicx}
\usepackage{textcomp}
\usepackage{xcolor}
\usepackage{mathrsfs}
\usepackage{psfrag}
\usepackage{cancel}

\makeatletter
\let\old@ssect\@ssect % Store how ifacconf defines \@ssect
\usepackage{natbib}
\usepackage{hyperref}
\hypersetup{colorlinks,linkcolor={blue},citecolor={blue},urlcolor={red}}
\def\@ssect#1#2#3#4#5#6{%
  \NR@gettitle{#6}% Insert key \nameref title grab
  \old@ssect{#1}{#2}{#3}{#4}{#5}{#6}% Restore ifacconf's \@ssect
}
\makeatother

\def\mc{\mathcal}

\newtheorem{definition}{Definition}
\newtheorem{theorem}{Theorem}

\newtheorem{remark}{Remark}

\begin{document}
\sloppy
\begin{frontmatter}
\title{\LARGE \bf Negative Imaginary State Feedback Equivalence for a Class of Nonlinear Systems\thanksref{footnoteinfo}}

\thanks[footnoteinfo]{This work was supported by the Australian Research Council under grant DP190102158.}

\author[First]{Kanghong Shi},
\author[First]{Ian R. Petersen},
\author[First]{Igor G. Vladimirov}

\address[First]{K. Shi, I. R. Petersen and I. G. Vladimirov are with the School of Engineering, College of Engineering and Computer Science, Australian National University, Canberra, ACT 2601, Australia.
        {\tt kanghong.shi@anu.edu.au}, {\tt ian.petersen@anu.edu.au}, {\tt igor.vladimirov@anu.edu.au}.}

%\author{Kanghong Shi,$\quad$Ian R. Petersen, \IEEEmembership{Fellow, IEEE},$\quad$and$\quad$Igor G. Vladimirov % <-this % stops a space
%\thanks{This work was supported by the Australian Research Council under grant DP190102158.}% <-this % stops a space
%\thanks{K. Shi, I. R. Petersen and I. G. Vladimirov are with the School of Engineering, College of Engineering and Computer Science, Australian National University, Canberra, Acton, ACT 2601, Australia.
%        {\tt kanghong.shi@anu.edu.au}, {\tt ian.petersen@anu.edu.au}, {\tt igor.vladimirov@anu.edu.au}.}%
%}

\maketitle
\thispagestyle{plain}
\pagestyle{plain}

\begin{abstract}
\input{Abstract.tex}

\end{abstract}

\begin{keyword}
nonlinear negative imaginary systems, state feedback equivalence, state feedback stabilization, robust control.
\end{keyword}

\end{frontmatter}

%%%%%%%%%%%%%%%%%%%%%%% Sections %%%%%%%%%%%%%%%%%%%%%%%%%%%
\input{Introduction.tex}

\input{Notation.tex}
\input{Preliminaries.tex}

\input{Feedback_equivalence_nonlinear_NI_with_pd_sf.tex}

\input{Conclusion.tex}

%\addtolength{\textheight}{-12cm}
%\input{Acknowledgment.tex}
%\bibliographystyle{ifac}

\input{main.bbl}
\end{document}

%% file: Abstract.tex
In this paper, we investigate the necessary and sufficient conditions under which a class of nonlinear systems are state feedback equivalent to nonlinear negative imaginary (NI) systems with positive definite storage functions. The nonlinear systems of interest have a normal form of relative degree less than or equal to two. The nonlinearity of the system is restricted with respect to a subset of the state variables, which are the state variables that have external dynamics. Under mild assumptions, such systems are state feedback equivalent to nonlinear NI systems and nonlinear output strictly negative imaginary (OSNI) systems if and only if they are weakly minimum phase. Such a state feedback control approach can also asymptotically stabilize the systems in question against nonlinear OSNI system uncertainties. A numerical example is provided to show the process of the state feedback equivalence control and stabilization of uncertain systems.

%% file: Introduction.tex
\section{INTRODUCTION}
Negative imaginary (NI) systems theory was introduced in \cite{lanzon2008stability} and \cite{petersen2010feedback}, to address the robust control problem for flexible structures \cite{preumont2018vibration,halim2001spatial,pota2002resonant}. Such systems have highly resonant dynamics, for which the commonly used negative velocity feedback control \cite{brogliato2007dissipative} may not be suitable. NI systems theory provides an alternative approach, which uses positive position feedback control. From this perspective, NI systems theory can be regarded as a complement to the positive real (PR) systems theory \cite{brogliato2007dissipative,khalil2002nonlinear}. Typical mechanical NI systems are systems with colocated force actuators and position sensors. NI systems can have relative degree zero, one and two while PR systems can only have relative degree zero and one. This enables NI systems theory to be applied to a broader class of systems in comparison to PR systems theory. NI systems theory has attracted attention from many control theorists since it was introduced in 2008  \cite{xiong2010negative,song2012negative,mabrok2014generalizing,wang2015robust,bhowmick2017lti}. It has been applied in many fields including nano-positioning control \cite{mabrok2013spectral,das2014mimo,das2014resonant,das2015multivariable} and the control of lightly damped structures \cite{cai2010stability,rahman2015design,bhikkaji2011negative}, etc.

Roughly speaking, a square real-rational proper transfer matrix $F(s)$ is said to be NI if it is stable and $j(F(j\omega)-F(j\omega)^*)\geq 0$ for all frequencies $\omega \geq 0$. Under mild assumptions, the positive feedback interconnection of an NI system $F(s)$ and a strictly negative imaginary (SNI) system $G(s)$ is internally stable if and only if the DC loop gain has all its eigenvalues less than unity; i.e., $\lambda_{max}(F(0)G(0))<1$.

NI systems theory was extended to nonlinear systems in \cite{ghallab2018extending,shi2021robust}. Roughly speaking, a system is said to be nonlinear NI if it has a positive definite storage function $V(x)$ such that $\dot V(x)\leq u^T\dot y$ for all $t\geq 0$, where $x$, $u$ and $y$ are the state, input and output of the system, respectively. To allow for systems with poles at the origin, the nonlinear NI systems definition is generalized in \cite{shi2020output}, which only requires positive semidefiniteness of the storage function. Also introduced in \cite{shi2021robust} and \cite{shi2020output} is the notion of nonlinear output strictly negative imaginary (OSNI) systems (see \cite{bhowmick2017lti} and \cite{bhowmickoutput} for the definition of linear OSNI systems). Under reasonable assumptions, the interconnection of a nonlinear NI system and a nonlinear OSNI system is asymptotically stable.

Many papers have investigated the problem of making a system of relative degree one passive or PR using state feedback control \cite{kokotovic1989positive,saberi1990global,byrnes1991passivity,byrnes1991asymptotic,santosuosso1997passivity,lin1995feedback,jiang1996passification}. These passivity and PR state feedback equivalence results are useful in stabilizing systems of relative degree one, due to the effectiveness of passivity and PR systems theory. The significance of such state feedback passivity equivalence lies not only in generalizing the feedback linearization method (see \cite{khalil2002nonlinear,isidori1995nonlinear}), but also in enabling the passivity theory to be applied to a broader class of systems, since feedback systems analysis and design is comparatively simple and intuitive for passive systems \cite{byrnes1991passivity}. However, due to the natural constraints of passive systems, these results are only applicable to systems of relative degree one. This rules out a wide variety of systems which have output entries of relative degree two.

As NI systems theory can deal with systems with relative degree zero, one and two, one may consider investigating the state feedback equivalence problem using NI systems theory; i.e., making a system NI using state feedback control. In \cite{shi2021negative}, the NI state feedback equivalence problem is investigated for linear systems of relative degree one and relative degree two separately. This result is then generalized in the paper \cite{shi2021necessary}, which considers the case that a system can have both relative degree one and relative degree two output entries. \cite{shi2021necessary} provides the necessary and sufficient conditions under which a linear system is state feedback equivalent to an NI, OSNI or strongly strictly negative imaginary (SSNI) system. Such an NI state feedback equivalence approach can be used to asymptotically stabilize systems with SNI uncertainties.

Considering the nonlinear nature of most control systems, \cite{shi2022making} addressed the problem of making affine nonlinear systems nonlinear NI using state feedback control. In addition, for a system that can be made nonlinear NI, if its internal dynamics is input-to-state stable (ISS) (see \cite{sontag1989smooth,sontag2008input} for ISS systems), then there exists a state feedback control law that stabilizes the system. The reason that the ISS condition is needed is because the approach in \cite{shi2022making} only makes the resulting system nonlinear NI with a positive semidefinite storage function.

This paper aims to complement the results in \cite{shi2022making} by making a system nonlinear NI with a positive definite storage function. The motivation for guaranteeing the positive definiteness of the storage function of the resulting system is to make it easy to construct a positive definite closed-loop Lyapunov function for feedback controller synthesis (see for example \cite{khalil2002nonlinear} for Lyapunov's direct method). Many of the passivity feedback equivalence papers referenced above also require the positive definiteness of the storage function of the resulting systems in order to achieve stability (see for example \cite{byrnes1991passivity}). With the help of a positive definite storage function for the resulting closed-loop system, the ISS condition is no longer required. The trade-off, on the other hand, is that internal dynamics (see \cite{isidori1995nonlinear,khalil2002nonlinear,byrnes1991asymptotic} for internal dynamics) of the original system in question should have a constrained nonlinearity. From a different perspective, the present work is a generalization of the linear results in \cite{shi2021necessary}, as the system in question in \cite{shi2021necessary} belongs to a subset of the class of systems investigated in the present paper. To avoid distracting the readers from the main results on state feedback equivalence, we introduce the system in question in its normal form. The readers are referred to \cite{shi2022making} and references therein for the transformation from a general nonlinear system to its normal form.

To be specific, this paper provides the necessary and sufficient conditions under which the system in question is state feedback equivalent to a nonlinear NI system or a nonlinear OSNI system. Formulas for the corresponding state feedback controller are given. Conditions are also provided for stabilization when the system in question has nonlinear OSNI uncertainty. The process of making a system nonlinear NI and stabilizing it against nonlinear OSNI uncertainty is illustrated later in a numerical example. The contribution of this paper is in solving the nonlinear NI state feedback equivalence problem, which broadens the class of systems to which nonlinear NI systems theory is applicable. Also, this paper provides a controller synthesis approach for systems with nonlinear OSNI uncertainties.

This paper is organized as follows: Section \ref{sec:pre} reviews the definitions of nonlinear NI and OSNI systems. Section \ref{sec:state feedback equivalence} provides necessary and sufficient conditions under which the system in question can be made nonlinear NI or nonlinear OSNI. Section \ref{sec:stabilization} applies the results presented in Section \ref{sec:state feedback equivalence} to address a stabilization problem for systems with nonlinear OSNI uncertainties. Section \ref{sec:example} illustrates the process of making a system nonlinear NI and stabilizing an uncertain system with a numerical example. A conclusion is given in Section \ref{sec:conclusion}.

%% file: Notation.tex
Notation: The notation in this paper is standard. $\mathbb R$ denotes the field of real numbers. $\mathbb R^{m\times n}$ denotes the space of real matrices of dimension $m\times n$. $A^T$ denotes the transpose of a matrix $A$. $spec(A)$ denotes the spectrum of $A$. $\lambda_{max}(A)$ denotes the largest eigenvalue of a matrix $A$ with real spectrum. $\|\cdot\|$ denotes the standard Euclidean norm. $C^k$ represents the class of $k$-time continuously differentiable functions. $CLHP$ is the closed left half-plane of the complex plane.

%% file: Preliminaries.tex
\section{PRELIMINARIES}\label{sec:pre}

Consider the following general nonlinear system:
\begin{subequations}\label{eq:general system}
\begin{align}
\Sigma: \quad   \dot x=&\ f(x,u),\label{eq:general system state}\\
    y=&\ h(x),\label{eq:general system output}
\end{align}	
\end{subequations}
where $x\in \mathbb R^n$, $u\in \mathbb R^p$ and $y\in \mathbb R^p$ are the state, input and output of the system. The function $f:\mathbb R^n \times \mathbb R^p \to \mathbb R^n$ is globally Lipschitz and $h:\mathbb R^n \to \mathbb R^p$. Here, $f,h$ are of class $C^\infty$. We suppose that the vector field $f$ has at least one equilibrium. Then without loss of generality, we can assume $f(0,0)=0$ and $h(0)=0$ after a coordinate shift.
\begin{definition}[Nonlinear NI Systems]\label{def:nonlinear NI}\cite{shi2020output,ghallab2018extending}
The system (\ref{eq:general system}) is said to be a nonlinear NI system if there exists a positive semidefinite storage function $V:\mathbb R^n\to \mathbb R$ of class $C^1$ such that
\begin{equation}\label{eq:NI MIMO definition inequality}
    \dot V(x(t))\leq u(t)^T\dot y(t)
\end{equation}
for all $t\geq 0$.
\end{definition}

\begin{definition}[Nonlinear OSNI Systems]\label{def:nonlinear OSNI}\cite{shi2020output}
The system (\ref{eq:general system}) is said to be a nonlinear OSNI system if there exists a positive semidefinite storage function $V:\mathbb R^n\to\mathbb R$ of class $C^1$ and a scalar $\epsilon>0$ such that
\begin{equation}\label{eq:dissipativity of OSNI}
    \dot V(x(t))\leq u(t)^T\dot { y}(t) -\epsilon \left\|\dot {y}(t)\right\|^2
\end{equation}
for all $t\geq 0$. In this case, we also say that system (\ref{eq:general system}) is nonlinear OSNI with degree of output strictness $\epsilon$.
\end{definition}

\begin{definition}(Lyapunov Stability) \cite{bernstein2009matrix}
\label{def:LS}
A square matrix $A$ is said to be Lyapunov stable if $spec(A)\subset CLHP$ and every purely imaginary eigenvalue of $A$ is semisimple.
\end{definition}

%% file: Feedback_equivalence_nonlinear_NI_with_pd_sf.tex
\section{State Feedback Equivalence to a Nonlinear NI System}\label{sec:state feedback equivalence}
We consider the following system
\begin{subequations}\label{eq:normal form}
\begin{align}
\dot z =&\ A_{11}z+p(y),\label{eq:normal form a}\\
\dot \xi_1=&\ u_1,\\
\dot \xi_2=&\ \xi_3,\\
\dot \xi_3=&\ u_2,\\
y=& \left[\begin{matrix}	\xi_1 \\ \xi_2 \end{matrix}
\right],
\end{align}
\end{subequations}
where $u= \left[\begin{matrix}	u_1^T & u_2^T \end{matrix}
\right]^T\in \mathbb R^p$ is the system input, $y \in \mathbb R^p$ is the system output and $\left[\begin{matrix}z^T & \xi \end{matrix}
\right]^T$ is the system state. Here, $z \in \mathbb R^{m}$ is a complementary state variable. $\xi=\left[\begin{matrix}	\xi_1^T & \xi_2^T & \xi_3^T  \end{matrix}
\right]^T$, where $\xi_1\in \mathbb R^{p_1}$ and $\xi_2,\xi_3 \in \mathbb R^{p_2}$ ($p_2 := p-p_1$). $p(0)=0$.

\begin{definition}
A system of the form (\ref{eq:general system}) is said to be state feedback equivalent to a nonlinear NI (OSNI) system with a positive definite storage function if
there exists a state feedback control law
\begin{equation}
u = k(x)+v,	
\end{equation}
where $v\in \mathbb R^p$ is the new input, such that the system with the input $v$ and the output $y$ is a nonlinear NI (OSNI) system with a positive definite storage function.
\end{definition}

We provide necessary and sufficient conditions for the nonlinear NI state feedback equivalence of the system (\ref{eq:general system}). Note that for the system (\ref{eq:general system}), it is said to have \emph{weakly minimum phase} if $A_{11}$ is Lyapunov stable (see \cite{khalil2002nonlinear,isidori1995nonlinear,byrnes1991asymptotic}).

\begin{theorem}\label{theorem:state feedback equivalence NNI}
Suppose the system (\ref{eq:normal form}) satisfies $\det A_{11} \neq 0$. Then it is state feedback equivalent to a nonlinear NI (OSNI) system with a positive definite storage function if and only if $A_{11}$ is Lyapunov stable.
\end{theorem}

\begin{pf}
\textbf{Sufficiency}. Given that $\dot z = A_{11}z$ is Lyapunov stable, there exists a storage function $V_1(z)>0$ under which for any $z$ we have that (see \cite{khalil2002nonlinear})
\begin{equation}\label{eq:weakly minimum phase}
	\frac{\partial V_1(z)}{\partial z}A_{11}z \leq 0.
\end{equation}
Let us define a function $V(z,\xi)$ as
\begin{equation}\label{eq:storage function}
V(z,\xi)=V_1(\alpha)+V_2(y)+\frac{1}{2}\xi_3^T\xi_3,	
\end{equation}
where $V_1(\cdot)$ satisfies (\ref{eq:weakly minimum phase}) and $V_2(y)$ can be any positive definite function. Here, a new vector $\alpha$ is defined as
\begin{equation}\label{eq:alpha def}
\alpha = z+A_{11}^{-1}p(y).	
\end{equation}
In this case, $V(z,\xi)$ is a positive definite function. Consider the following inequality, which will be used later. According to (\ref{eq:weakly minimum phase}), we have that
\begin{equation}
\frac{\partial V(z,\xi)}{\partial z}[A_{11}z+p(y)] = \frac{\partial V_1(\alpha)}{\partial z}A_{11}\alpha = \frac{\partial V_1(\alpha)}{\partial \alpha}A_{11}\alpha \leq 0.
\end{equation}
Let us consider the following change of input. We change the input entries $u_1$ to be
\begin{equation}\label{eq:u_1 nonlinear}
u_1=v_1-\left(\frac{\partial V(z,\xi)}{\partial \xi_1}\right)^T,
\end{equation}
and we change the input entries $u_2$ to be
\begin{equation}\label{eq:u_2 nonlinear}
u_2=v_2-\left(\frac{\partial V(z,\xi)}{\partial \xi_2}\right)^T-\lambda \xi_3,
\end{equation}
where $\lambda \geq 0$ is a scalar. Here, $v_1\in \mathbb R^{p_1}$ and $v_2\in \mathbb R^{p_2}$, and $v = \left[\begin{matrix}v_1^T & v_2^T \end{matrix}
\right]^T$ is the new system input. The  resulting new system has the model:
\begin{subequations}\label{eq:resulting system}
\begin{align}
\dot z =&\ A_{11}z+p(y),\\
\dot \xi_1=&\ v_1-\left(\frac{\partial V(z,\xi)}{\partial \xi_1}\right)^T,\\
\dot \xi_2=&\ \xi_3,\\
\dot \xi_3=&\ v_2-\left(\frac{\partial V(z,\xi)}{\partial \xi_2}\right)^T-\lambda \xi_3,\\
y=& \left[\begin{matrix}	\xi_1 \\ \xi_2 \end{matrix}
\right].
\end{align}
\end{subequations}
Now, let us verify the nonlinear NI property of the new system (\ref{eq:resulting system}). We have that
\begin{align*}
\dot V&(z,\xi)-v^T\dot y\notag\\
=& \frac{\partial V(z,\xi)}{\partial z}\dot z+	\frac{\partial V(z,\xi)}{\partial \xi}\dot \xi-v^T\dot y\notag\\
=&\frac{\partial V(z,\xi)}{\partial z}[A_{11}z+p(y)]+\frac{\partial V(z,\xi)}{\partial \xi_1}\dot \xi_1+\frac{\partial V(z,\xi)}{\partial \xi_2}\dot \xi_2\notag\\
&+\frac{\partial V(z,\xi)}{\partial \xi_3}\dot \xi_3-v_1^T\dot \xi_1-v_2^T\dot \xi_2\notag\\
\leq & -\left(v_1^T-\frac{\partial V(z,\xi)}{\partial \xi_1}\right) \dot \xi_1 -\left(v_2^T-\frac{\partial V(z,\xi)}{\partial \xi_2}\right)\dot \xi_2\notag\\
&+ \frac{\partial V(z,\xi)}{\partial \xi_3}\dot \xi_3 \notag\\
\leq & -\left(v_1^T-\frac{\partial V(z,\xi)}{\partial \xi_1}\right)\left(v_1-\left(\frac{\partial V(z,\xi)}{\partial \xi_1}\right)^T\right)\notag\\
&-\left(v_2^T-\frac{\partial V(z,\xi)}{\partial \xi_2}\right)\xi_3+\xi_3^T\left(v_2-\left(\frac{\partial V(z,\xi)}{\partial \xi_2}\right)^T-\lambda \xi_3\right)\notag \\
=& -\left(v_1^T-\frac{\partial V(z,\xi)}{\partial \xi_1}\right)\left(v_1-\left(\frac{\partial V(z,\xi)}{\partial \xi_1}\right)^T\right)-\lambda \xi_3^T\xi_3\notag\\
=&-\|\dot\xi_1\|^2-\lambda \|\dot \xi_2\|^2 \leq -\epsilon\|\dot y\|^2 \leq  0,\notag
\end{align*}
where $\epsilon=\min \{1,\lambda\}$. For all $\lambda\geq 0$, the new system is nonlinear NI. For $\lambda>0$, we have that $\epsilon>0$. Hence, the new system is nonlinear OSNI. This completes the sufficiency part of the proof.

\textbf{Necessity}. Note that nonlinear OSNI systems belong to the class of nonlinear NI systems. Therefore, suppose the system is state feedback equivalent to an NI system with a positive definite storage function. This implies that there is a state feedback control law
\begin{equation*}
u = \left[\begin{matrix}	u_1 \\ u_2 \end{matrix}
\right]=\left[\begin{matrix}	k_1(z,\xi)+v_1 \\ k_2(z,\xi)+v_2 \end{matrix}
\right],
\end{equation*}
such that the resulting system
\begin{align*}
\dot z =&\ A_{11}z+p(y),\\
\dot \xi_1=&\ v_1+ k_1(z,\xi),\\
\dot \xi_2=&\ \xi_3,\\
\dot \xi_3 =&\ v_2+k_2(z,\xi),\\
y=& \left[\begin{matrix}	\xi_1 \\ \xi_2 \end{matrix}
\right]
\end{align*}
with the new input $v = \left[\begin{matrix}v_1^T & v_2^T \end{matrix}
\right]^T$ is a nonlinear NI system with a positive definite storage function $V(z,\xi)$ that satisfies
\begin{equation}\label{eq:wmp}
\frac{\partial V(z,\xi)}{\partial z}\dot z + \frac{\partial V(z,\xi)}{\partial \xi} \dot \xi \leq v^T \dot y.
\end{equation}
Choose $v$ such that $y$ stays identically at zero. In this case, $\xi=0$, $\dot \xi =0$ and $p(y)=p(0)=0$. Then $V(z,0)$, which is a positive definite storage function of $z$, shows the Lyapunov stability of the system $\dot z = A_{11}z$ according to (\ref{eq:wmp}).
\end{pf}
\begin{remark}
The system of interest in Theorem \ref{theorem:state feedback equivalence NNI} is a particular normal form of a general system (\ref{eq:general system}) in the case that the system (\ref{eq:general system}) has relative degree less than or equal to two. Systems with such a relative degree condition have been investigated in \cite{shi2021necessary} and \cite{shi2022making}. The state-space model (\ref{eq:normal form}) contains a broader class of systems than that investigated in \cite{shi2021necessary} because \cite{shi2021necessary} only considers linear systems while in (\ref{eq:normal form a}), linearity is only required in the term $A_{11}z$. Compared to the system of interest in \cite{shi2022making}, Theorem \ref{theorem:state feedback equivalence NNI} sacrifices some generality in the system model but achieves a stronger property for the resulting system. That is, the resulting system achieved via state feedback is not only nonlinear NI (OSNI), but also has a positive definite storage function. As one might notice, the system considered in Theorem \ref{theorem:state feedback equivalence NNI} is already in a special form, which frees us from presenting the transformation from the general system (\ref{eq:general system}) to the form (\ref{eq:normal form}). The transformation, which requires complex conditions and procedures, is not the main focus of this paper and is relatively separate from the main results on nonlinear NI state feedback equivalence. Briefly speaking, the transformation from (\ref{eq:general system}) to (\ref{eq:normal form}) requires feedback linearization (see \cite{khalil2002nonlinear}), where state transformation and change of input also need to be applied. Conditions and procedures are similar to what is presented in \cite{shi2021necessary} and \cite{shi2022making}. Also, note that we only consider systems with relative degree less than or equal to two because NI systems cannot have output entries with relative degree greater than two \cite{shi2021necessary}. However, allowing for systems with output entries of relative degree two already provides significant contribution because such systems cannot be dealt with by results in the passivity state feedback equivalence papers (for example \cite{byrnes1991passivity}).
\end{remark}

\begin{remark}
The system considered in Theorem \ref{theorem:state feedback equivalence NNI} is in a normal form where a change of input is already applied. A more general version of the normal form is
\begin{subequations}\label{eq:normal form general}
\begin{align}
\dot z =&\ A_{11}z+p(y),\\
\dot \xi_1=&\ j_1(z,\xi)+ l_1(z,\xi) \tilde u_1,\\
\dot \xi_2=&\ \xi_3,\\
\dot \xi_3=&\ j_2(z,\xi)+ l_2(z,\xi) \tilde u_2,\\
y=& \left[\begin{matrix}	\xi_1 \\ \xi_2 \end{matrix}
\right],
\end{align}
\end{subequations}	
where $\left[\begin{matrix}	l_1(z,\xi) \\ l_2(z,\xi) \end{matrix}
\right]$ is nonsingular for all $(z,\xi)$ in the region of interest, e.g., near the equilibrium point $(0,0)$ or globally. Then, the system (\ref{eq:normal form general}) can be transformed into the form (\ref{eq:normal form}) using the following change of input:
\begin{equation*}
\left[\begin{matrix}	\tilde u_1 \\ \tilde u_2 \end{matrix}
\right] = \left[\begin{matrix}	l_1(z,\xi) \\ l_2(z,\xi) \end{matrix}
\right]^{-1} \left[\begin{matrix}	u_1-j_1(z,\xi) \\ u_2-j_2(z,\xi) \end{matrix}
\right],
\end{equation*}
\end{remark}
where $u = \left[\begin{matrix}	u_1 \\ u_2 \end{matrix}
\right]$ is the new input.

\section{Stabilization of Uncertain Systems with nonlinear OSNI uncertainties}\label{sec:stabilization}
\begin{figure}[h!]
\centering
\psfrag{delta}{\hspace{0.1cm}$\mc H_\sigma$}
\psfrag{nominal}{\hspace{-0.05cm}\small Nominal}
\psfrag{plant}{\hspace{-0.2cm}\small Plant $\Sigma$}
\psfrag{controller}{\small Controller}
\psfrag{closed-loop}{\hspace{-0.07cm}\small Closed-Loop}
\psfrag{w}{$w$}
\psfrag{x}{$x$}
\psfrag{y}{$y$}
\psfrag{u}{$u$}
\psfrag{R_s}{\hspace{0.03cm}\small $\mc H_p$}
\psfrag{+}{\small$+$}
\includegraphics[width=8.5cm]{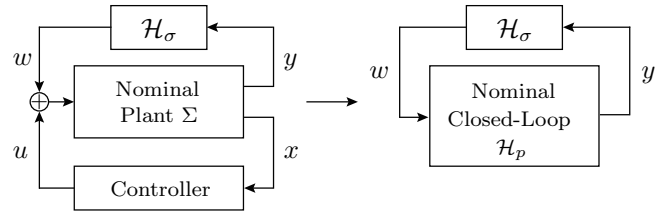}
\caption{A feedback control system. The plant uncertainty $\mc H_\sigma$ is a nonlinear OSNI system. Under some assumptions, we can find a state feedback controller such that the closed-loop system is asymptotically stable.}
\label{fig:controller synthesis}
\end{figure}
Consider a system $\Sigma$ of the form (\ref{eq:normal form}) having a system uncertainty $\mc H_\sigma$ that can be described as a nonlinear OSNI system. We show in the following that such an uncertain system can be asymptotically stabilized using a state feedback controller, as shown on the left-hand side (LHS) of Fig.~\ref{fig:controller synthesis}, that makes the nominal closed-loop system nonlinear OSNI. Suppose the uncertainty $\mc H_\sigma$ can be described by the following equations:
\begin{align*}
H_\sigma:\quad\dot x_\sigma =&\ f_\sigma(x_\sigma,u_\sigma),\\
y_\sigma =&\ h_\sigma(x_\sigma),
\end{align*}	
where $x_\sigma\in \mathbb R^{n_\sigma}$ is the state, $u_\sigma\in \mathbb R^p$ is the input, and $y_\sigma\in \mathbb R^p$ is the output, $f_\sigma:\mathbb R^{n_\sigma}\times \mathbb R^p \to \mathbb R^{n_\sigma}$ is a Lipschitz continuous function and $h_\sigma:\mathbb R^{n_\sigma} \to \mathbb R^p$ is a class $C^1$ function. Suppose the system has at least one equilibrium. Then without loss of generality, we can assume $f_\sigma(0,0)=0$ and $h_\sigma(0)=0$ after a possible coordinate shift.

The closed-loop interconnection of the nominal plant $\Sigma$ and the uncertainty $\mc H_\sigma$ can be described as follows:
\begin{subequations}\label{eq:interconnection}
\begin{align}
\dot z =&\ A_{11}z+p(y),\\
\dot \xi_1=&\ (u_1+w_1),\\
\dot \xi_2=&\ \xi_3,\\
\dot \xi_3=&\ (u_2+w_2),\\
y=& \left[\begin{matrix}	\xi_1 \\ \xi_2 \end{matrix}
\right],\\
\dot x_\sigma =&\ f_\sigma(x_\sigma,u_\sigma),\\
y_\sigma =&\ h_\sigma(x_\sigma),\\
w=&\ y_\sigma,\label{eq:interconnection h}\\
u_\sigma=&\ y,\label{eq:interconnection i}
\end{align}
\end{subequations}
where $w_1$ contains the first $p_1$ entries of the output $w$ of the uncertainty $\mc H_\sigma$ and $w_2$ contains the rest of the output entries; i.e., $w = \left[\begin{matrix}	w_1 \\ w_2 \end{matrix}
\right]$.

\begin{theorem}\label{theorem:stabilization}
Consider the interconnection of the uncertain system $\Sigma$ and the system uncertainty $\mc H_\sigma$ as described by (\ref{eq:interconnection}) and shown in Fig.~\ref{fig:controller synthesis}. Suppose $\det{A_{11}}\neq 0$ and $A_{11}$ is Lyapunov stable. Also, suppose the system $\mc H_\sigma$ is OSNI with the storage function $V_\sigma(x_\sigma)$. If the function, given by
\begin{equation}\label{eq:W}
W(z,\xi,x_\sigma)=V(z,\xi)+V_\sigma(x_\sigma)-h_\sigma(x_\sigma)^T \left[\begin{matrix}	\xi_1 \\ \xi_2 \end{matrix}
\right]
\end{equation}
with $V(z,\xi)$ defined in (\ref{eq:storage function}), is positive definite, then the system (\ref{eq:interconnection}) is asymptotically stabilized by the state feedback control laws
\begin{equation}\label{eq:u_1 uncertain}
u_1=-\left(\frac{\partial V(z,\xi)}{\partial \xi_1}\right)^T,
\end{equation}
and
\begin{equation}\label{eq:u_2 uncertain}
u_2=-\left(\frac{\partial V(z,\xi)}{\partial \xi_2}\right)^T-\lambda \xi_3,
\end{equation}
where $\lambda>0$ is a scalar.
\end{theorem}
\begin{pf}
After applying the state feedback control laws (\ref{eq:u_1 uncertain}) and (\ref{eq:u_2 uncertain}), the nominal plant $\Sigma$ becomes the nominal closed-loop system $\mc H_p$, which is shown on the right-hand side (RHS) of Fig.~\ref{fig:controller synthesis}. The input of the system $\mc H_p$ is the output $w$ of the uncertainty $\mc H_\sigma$. As is shown in Section \ref{sec:state feedback equivalence}, under the control laws (\ref{eq:u_1 uncertain}) and (\ref{eq:u_2 uncertain}), the resulting nominal closed-loop system $\mc H_p$ is a nonlinear OSNI system with input $w$. The corresponding positive definite storage function $V(z,\xi)$ satisfies
\begin{equation*}
	\dot V(z,\xi)\leq w^T \dot y-\epsilon\|\dot y\|^2,
\end{equation*}
where $\epsilon=\min\{1,\lambda\}$, as is shown in the proof of Theorem \ref{theorem:state feedback equivalence NNI}. Also, because $\mc H_\sigma$ is also a nonlinear OSNI system, its storage function $V_\sigma(x_\sigma)$ satisfies that
\begin{align*}
\dot V_\sigma(x_\sigma)\leq &\ u_\sigma^T\dot y_\sigma-\epsilon_\sigma \|\dot y_\sigma\|^2\notag\\
	=& \ y^T \dot w - \epsilon_\sigma \|\dot w\|^2,
\end{align*}
where the equality also uses the equations (\ref{eq:interconnection h}) and (\ref{eq:interconnection i}). Here, the scalar $\epsilon_\sigma>0$ quantifies the level of output strictness in $\mc H_\sigma$. Now, we use Lyapunov's direct method to prove the stability of the interconnection shown on the RHS of Fig.~\ref{fig:controller synthesis}. Let us consider $W(z,\xi,x_\sigma)$ given in (\ref{eq:W}) as the storage function of the interconnection. Taking the time derivative of $W(z,\xi,x_\sigma)$, we have
\begin{align}
\dot W&(z,\xi,x_\sigma)\notag\\
 =&\ \dot V(z,\xi)+\dot V(x_\sigma)-\dot h_\sigma(x_\sigma)^T\left[\begin{matrix}	\xi_1 \\ \xi_2 \end{matrix}
\right]-h_\sigma(x_\sigma)^T	\left[\begin{matrix}	\dot \xi_1 \\ \dot \xi_2 \end{matrix}
\right]\notag\\
\leq &\ w^T \dot y - \epsilon \|\dot y\|^2 +u_\sigma^T\dot y_\sigma - \epsilon_\sigma \|\dot y_\sigma\|^2 - \dot h_\sigma(x_\sigma)^T\left[\begin{matrix}	\xi_1 \\ \xi_2 \end{matrix}
\right]\notag\\
&-h_\sigma(x_\sigma)^T	\left[\begin{matrix}	\dot \xi_1 \\ \dot \xi_2 \end{matrix}
\right]\notag\\
=&\ w^T \dot y - \epsilon \|\dot y\|^2 +y^T\dot w - \epsilon_\sigma \|\dot w\|^2 - \dot w^T y-w^T	\dot y \notag\\
=& - \epsilon \|\dot y\|^2 - \epsilon_\sigma \|\dot w\|^2\notag\\
\leq &\ 0.\label{eq:dot W ineq}
\end{align}
We have that $\dot W(z,\xi,x_\sigma)\leq 0$ and $\dot W(z,\xi,x_\sigma)=0$ only if $\dot y=0$ and $\dot w=0$. We apply LaSalle's invariance principal in the following. Observing (\ref{eq:dot W ineq}), it is possible for $\dot W(z,\xi,x_\sigma)$ to remain at zero only if $\dot y=0$ and $\dot w=0$ hold over a time interval $[t_a,t_b]$ with $t_a<t_b$; i.e., $\dot y\equiv 0$ and $\dot w \equiv 0$. And $\dot y \equiv 0$ implies that $\dot \xi_1 \equiv 0$ and $\dot \xi_2 =\xi_3\equiv 0$. Consider (\ref{eq:resulting system}), with $v_1$ and $v_2$ replaced by $w_1$ and $w_2$, respectively. We have that
\begin{equation*}
\dot \xi_1 \equiv 0 \implies 	w_1 \equiv \left(\frac{\partial V(z,\xi)}{\partial \xi_1}\right)^T;
\end{equation*}
\begin{equation*}
\dot \xi_2 \equiv 0 \implies w_2\equiv \left(\frac{\partial V(z,\xi)}{\partial \xi_2}\right)^T.
\end{equation*}
Also, since $\dot w \equiv 0$, the system $\mc H_\sigma$ is in steady state. That is, given constant input $u = \left[\begin{matrix}	\xi_1 \\ \xi_2 \end{matrix}
\right]$, the system $\mc H_\sigma$ also has constant output $y_\sigma=\left[\begin{matrix}	\left(\frac{\partial V(z,\xi)}{\partial \xi_1}\right)^T \\ \left(\frac{\partial V(z,\xi)}{\partial \xi_2}\right)^T \end{matrix}
\right]$. Consider that
\begin{align*}
	\left[\begin{matrix}	\left(\frac{\partial V(z,\xi)}{\partial \xi_1}\right)^T \\ \left(\frac{\partial V(z,\xi)}{\partial \xi_2}\right)^T \end{matrix}
\right] =& \left(\frac{\partial V(z,\xi)}{\partial y}\right)^T\notag\\
 =& \left(\frac{\partial V_1(\alpha)}{\partial y}\right)^T+\left(\frac{\partial V_2(y)}{\partial y}\right)^T, 
\end{align*}
which is a function of $z$ and $y$. Also, since $\dot y \equiv 0$, we have that $\left(\frac{\partial V_2(y)}{\partial y}\right)^T$ is constant. Therefore, the term
\begin{align*}
	\frac{\partial V_1(\alpha)}{\partial y} =& \frac{\partial V_1(\alpha)}{\partial \alpha}\frac{\partial \alpha}{\partial y}=\frac{\partial V_1(\alpha)}{\partial \alpha}\left(\frac{\partial z}{\partial y}+A_{11}^{-1}\frac{\partial p(y)}{\partial y}\right)
\end{align*}
must be constant. Since $y$ is constant, then $z$ must also be constant; i.e., $\dot z = 0$. We prove in the following that under the situation that $\dot y = 0$ and $\dot z = 0$, we also have $z=0$ and $\xi=0$ if a suitable $V_2(y)$ is chosen. Suppose the steady state input-output relation of the system $\mc H_\sigma$ can be described by some function $\bar y = \kappa (\bar u)$, where $\bar y$ and $\bar u$ are the constant output and input in steady state, respectively. Then we can always add additional positive definite function $\tilde V_2(y)$ to $W(z,\xi,x_\sigma)$ such that the curve of $\left[\begin{matrix}	\left(\frac{\partial V(z,\xi)}{\partial \xi_1}\right)^T \\ \left(\frac{\partial V(z,\xi)}{\partial \xi_2}\right)^T \end{matrix}
\right]$ intersects with $\kappa (\left[\begin{matrix}	\xi_1 \\ \xi_2 \end{matrix}
\right])$ only at the origin. In this case, the interconnection shown on the RHS of Fig.~\ref{fig:controller synthesis} cannot remain in steady state unless $\xi_1\equiv 0$ and $\xi_2 \equiv 0$. Therefore, $\xi_3 = \dot \xi_2 = 0$ and $y = 0$. This implies that $p(y)=p(0)=0$ and $\dot z = A_{11}z$. Since $\dot z = 0$, we have that $z = A_{11}^{-1}\dot z=0$. Therefore, in the case that $\dot y$ and $\dot w$ are zero, the system state is already at the origin. Otherwise, $W(z,\xi,x_\sigma)$ will keep decreasing until $z,\xi,x_\sigma$ all become zero. This completes the proof.
\end{pf}

\section{Example}\label{sec:example}
In this section, we illustrate the process of making a system of the form (\ref{eq:normal form}) nonlinear NI with a positive definite storage function using state feedback control. Consider the following system:
\begin{subequations}\label{eq:Example}
\begin{align}
\dot z =&\ -z+\xi_1^2\xi_2,\\
\dot \xi_1=&\ u_1,\\
\dot \xi_2=&\ \xi_3,\\
\dot \xi_3=&\ u_2,\\
y=& \left[\begin{matrix}	\xi_1 \\ \xi_2 \end{matrix}
\right].
\end{align}
\end{subequations}
where $u= \left[\begin{matrix}	u_1 & u_2 \end{matrix}
\right]^T\in \mathbb R^2$ is the system input, $y \in \mathbb R^2$ is the system output and $\left[\begin{matrix}z & \xi_1 & \xi_2 & \xi_3 \end{matrix}
\right]^T$ is the system state. Here, $z,\xi_1,\xi_2,\xi_3,u_1,u_2\in \mathbb R$. The corresponding $A_{11}$ matrix is $-1$. According to Theorem \ref{theorem:state feedback equivalence NNI}, the system is state feedback equivalent to a nonlinear NI (OSNI) system with a positive definite storage function. We show in the following the corresponding state feedback control and we prove that the resulting system is nonlinear NI. We construct
\begin{equation*}
V_1(\alpha) = \alpha^2,	
\end{equation*}
where $\alpha$ is defined as
\begin{equation*}
\alpha = z + A_{11}^{-1}\xi_1^2\xi_2 = z-\xi_1^2\xi_2,
\end{equation*}
according to (\ref{eq:alpha def}). We also construct
\begin{equation*}
V_2(y) = \xi_1^{\frac{4}{3}}+\xi_2^2.	
\end{equation*}
Therefore, the function $V(z,\xi)$ as given in (\ref{eq:storage function}) is
\begin{align}
V(z,\xi) =&\ (z-\xi_1^2\xi_2)^2+\xi_1^{\frac{4}{3}}+\xi_2^2+\frac{1}{2}\xi_3^2\notag\\
 =&\ z^2-2z\xi_1^2\xi_2+\xi_1^4\xi_2^2+\xi_1^{\frac{4}{3}}+\xi_2^2+\frac{1}{2}\xi_3^2.\label{eq:V in example}
\end{align}
We apply the control laws
\begin{equation}\label{eq:u_1 example}
u_1=v_1-\left(\frac{\partial V(z,\xi)}{\partial \xi_1}\right)^T=v_1+4z\xi_1\xi_2-4\xi_1^3\xi_2^2-\frac{4}{3}\xi_1^{\frac{1}{3}},
\end{equation}
and
\begin{equation}\label{eq:u_2 example}
u_2=v_2-\left(\frac{\partial V(z,\xi)}{\partial \xi_2}\right)^T-\lambda \xi_3=v_2+2z\xi_1^2-2\xi_1^4\xi_2-2\xi_2-\xi_3,
\end{equation}
where $v = \left[\begin{matrix}	v_1 & v_2 \end{matrix}
\right]^T$ is the new input. We choose $\lambda=1$ in (\ref{eq:u_2 example}). The resulting system now becomes
\begin{subequations}\label{eq:Example resulting}
\begin{align}
\dot z =&\ -z+\xi_1^2\xi_2,\\
\dot \xi_1=&\ v_1+4z\xi_1\xi_2-4\xi_1^3\xi_2^2-\frac{4}{3}\xi_1^{\frac{1}{3}},\\
\dot \xi_2=&\ \xi_3,\\
\dot \xi_3=&\ v_2+2z\xi_1^2-2\xi_1^4\xi_2-2\xi_2-\xi_3,\\
y=& \left[\begin{matrix}	\xi_1 \\ \xi_2 \end{matrix}
\right].
\end{align}
\end{subequations}
We prove that this system is nonlinear NI (OSNI) with the positive definite storage function (\ref{eq:V in example}). Take the derivative of $V(z,\xi)$, we get
\begin{align*}
\dot V(z,\xi) =&\ 2z\dot z-2\dot z \xi_1^2\xi_2-4z\xi_1\xi_2\dot \xi_1 - 2z\xi_1^2\dot \xi_2+4\xi_1^3\xi_2^2\dot \xi_1\notag\\
&+2\xi_1^4\xi_2\dot \xi_2+\frac{4}{3}\xi_1^{\frac{1}{3}}\dot \xi_1 + 2\xi_2\dot \xi_2+\xi_3\dot \xi_3\notag\\
=&\ (2z-2\xi_1^2\xi_2)\dot z + (4\xi_1^3\xi_2^2-4z\xi_1\xi_2+\frac{4}{3}\xi_1^{\frac{1}{3}})\dot \xi_1\notag\\
& + (2\xi_1^4\xi_2+2\xi_2-2z\xi_1^2)\dot \xi_2 + \xi_3\dot \xi_3\notag\\
=& -2\dot z^2+(v_1-\dot \xi_1)\dot \xi_1+\xi_3(v_2-\xi_3-\dot \xi_3)+\xi_3\dot \xi_3\notag\\
=& -2\dot z^2 + v_1\dot \xi_1-\dot \xi_1^2+v_2\xi_3-\xi_3^2\notag\\
=& -2\dot z^2 + v_1\dot \xi_1+v_2\dot \xi_2-\dot \xi_1^2-\dot \xi_2^2\notag\\
\leq &\ v^T\dot y-\|\dot y\|^2.
\end{align*}
Therefore, the system (\ref{eq:Example resulting}) is nonlinear OSNI and hence also nonlinear NI.

Now we consider an uncertainty for the system (\ref{eq:Example}). Suppose the uncertainty $\mc H_\sigma$ has the following model:
\begin{subequations}\label{eq:example NI uncertainty}
\begin{align}
\dot x_{\sigma 1} =& -x_{\sigma 1}^3+u_{\sigma 1},\\
\dot x_{\sigma 2} =& -x_{\sigma 2}+u_{\sigma 2},\\
y_{u} =& \left[\begin{matrix}x_{\sigma 1}\\x_{\sigma 2}
\end{matrix}
\right],
\end{align}
\end{subequations}
where $x_\sigma = [x_{\sigma 1}\ x_{\sigma 2}]^T$, $u_\sigma = [u_{\sigma 1}\ u_{\sigma 2}]^T$ and $y_\sigma = [x_{\sigma 1}\ x_{\sigma 2}]^T$ are the state, input and output of the system, respectively. Here, $x_{\sigma 1}, x_{\sigma 2}, u_{\sigma 1}, u_{\sigma 2}\in \mathbb R$.
The system (\ref{eq:example NI uncertainty}) is nonlinear OSNI with the positive definite storage function
\begin{equation*}
V_\sigma(x_\sigma) = \frac{1}{4}x_{\sigma 1}^4+\frac{1}{2}x_{\sigma 2}^2,
\end{equation*}
which satisfies the nonlinear OSNI property since
\begin{align*}
\dot V(x_\sigma)=&x_{\sigma 1}^3\dot x_{\sigma 1}+x_{\sigma 2}\dot x_{\sigma 2}\notag\\
=& (u_{\sigma 1}-\dot x_{\sigma 1})\dot x_{\sigma 1}+(u_{\sigma 2}-\dot x_{\sigma 2})\dot x_{\sigma 2}\notag\\
=& u_{\sigma 1}\dot x_{\sigma 1}+u_{\sigma 2}\dot x_{\sigma 2}-\dot x_{\sigma 1}^2-\dot x_{\sigma 2}^2\notag\\
= & u_\sigma^T \dot y_\sigma - \|\dot y_\sigma\|^2.
\end{align*}
Suppose the uncertain system has a structure shown on the LHS of Fig.~\ref{fig:controller synthesis}, we apply the state feedback control laws (\ref{eq:u_1 example}) and (\ref{eq:u_2 example}) to the system, with $w_1$ and $w_2$ replacing $v_1$ and $v_2$, respectively. Then the entire system becomes the system shown on the RHS of Fig.~\ref{fig:controller synthesis}. We show in the following that this interconnection is asymptotically stable. We construct the storage function of the interconnection using the formula given in (\ref{eq:W}):
\begin{equation*}
W(z,\xi,x_\sigma) = 	V(z,\xi)+ V_\sigma(x_\sigma) - \xi_1x_{\sigma 1}-\xi_2x_{\sigma 2}.
\end{equation*}
It can be verified that this storage function is positive definite. Then
\begin{align*}
\dot W& (z,\xi,x_\sigma)\notag\\
=&\ \dot V(z,\xi) + \dot V_\sigma(x_\sigma)-\dot \xi_1x_{\sigma 1}-\dot \xi_2 x_{\sigma 2} - \xi_1 \dot x_{\sigma 1}-\xi_2 
\dot x_{\sigma 2}\notag\\
=&-2\dot z^2 + w_1\dot \xi_1+w_2\dot \xi_2-\dot \xi_1^2-\dot \xi_2^2 + u_{\sigma 1}\dot x_{\sigma 1}+u_{\sigma 2}\dot x_{\sigma 2}\notag\\
&-\dot x_{\sigma 1}^2-\dot x_{\sigma 2}^2-\dot \xi_1x_{\sigma 1}-\dot \xi_2 x_{\sigma 2} - \xi_1 \dot x_{\sigma 1}-\xi_2 
\dot x_{\sigma 2}\notag\\
=&-2\dot z^2 + x_{\sigma 1}\dot \xi_1+x_{\sigma 2}\dot \xi_2-\dot \xi_1^2-\dot \xi_2^2 + \xi_1\dot x_{\sigma 1}+\xi_2\dot x_{\sigma 2}\notag\\
&-\dot x_{\sigma 1}^2-\dot x_{\sigma 2}^2-\dot \xi_1x_{\sigma 1}-\dot \xi_2 x_{\sigma 2} - \xi_1 \dot x_{\sigma 1}-\xi_2 
\dot x_{\sigma 2}\notag\\
=& -2\dot z^2 -\dot \xi_1^2-\dot \xi_2^2-\dot x_{\sigma 1}^2-\dot x_{\sigma 2}^2\notag\\
\leq &\ 0.
\end{align*}
Using LaSalle's invariance principal, $\dot W(z,\xi,x_\sigma)$ remains at zero only if $\dot z,\dot \xi_1,\dot \xi_2,\dot x_{\sigma 1},\dot x_{\sigma 2}\equiv 0$. In this case, we have that
\begin{align*}
\dot z =& -z+\xi_1^2\xi_2\equiv 0,\\
\dot \xi_1=&\ v_1+4z\xi_1\xi_2-4\xi_1^3\xi_2^2-\frac{4}{3}\xi_1^{\frac{1}{3}}\equiv 0,\\
\dot \xi_2=&\ \xi_3\equiv 0 \implies \dot \xi_3= v_2+2z\xi_1^2-2\xi_1^4\xi_2-2\xi_2-\xi_3\equiv 0,\\
\dot x_{\sigma 1} =& -x_{\sigma 1}^3+u_{\sigma 1}\equiv 0,\\
\dot x_{\sigma 2} =& -x_{\sigma 2}+u_{\sigma 2}\equiv 0.
\end{align*}
Solving the equations, we have $z,\xi_1,\xi_2,\xi_3,x_{\sigma 1},x_{\sigma 2}=0$. Otherwise, $W(z,\xi,x_\sigma)$ will keep decreasing until $z,\xi,x_\sigma$ all become zero. Therefore, the interconnection is asymptotically stable.

\begin{figure}[h!]
\centering
\psfrag{State}{State}
\psfrag{time (s)}{Time (s)}
\psfrag{States of the Nominal Plant of the Uncertain System}{\hspace{0.3cm} State Trajectories of the Nominal Plant}
\psfrag{z}{\scriptsize$z$}
\psfrag{x1}{\scriptsize$\xi_1$}
\psfrag{x2}{\scriptsize$\xi_2$}
\psfrag{x3}{\scriptsize$\xi_3$}
\psfrag{0}{\small$0$}
\psfrag{2}{\small$2$}
\psfrag{4}{\small$4$}
\psfrag{6}{\small$6$}
\psfrag{8}{\small$8$}
\psfrag{10}{\small$10$}
\psfrag{1}{\small$1$}
\psfrag{3}{\small$3$}
\psfrag{-1}{\hspace{-0.16cm}\small$-1$}
\includegraphics[width=9cm]{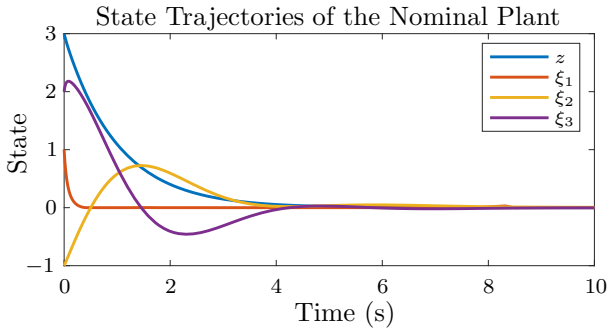}
\caption{State trajectories of the uncertain system (\ref{eq:Example}). The uncertainty of the system is described by (\ref{eq:example NI uncertainty}) and stability is achieved using the state feedback control (\ref{eq:u_1 example}) and (\ref{eq:u_2 example}) constructed according to Theorem \ref{theorem:stabilization}. Starting from nonzero initial values, the state variables of the nominal closed-loop system converge to zero, despite the presence of nonlinear OSNI plant uncertainty (\ref{eq:example NI uncertainty}).}
\label{fig:simulation}
\end{figure}

We also verify that the uncertain system is asymptotically stabilized via simulation. Let the initial state of the nominal plant be $x(0) = \left[\begin{matrix}z(0)& \xi_1(0)&\xi_2(0)&\xi_3(0)\end{matrix}\right]=\left[\begin{matrix}3& 1&-1&2\end{matrix}\right]$ and the initial state of the uncertainty be zero. It is shown in Fig.~\ref{fig:simulation} that despite the presence of the uncertainty, the system (\ref{eq:Example}) is asymptotically stabilized by the state feedback control (\ref{eq:u_1 example}) and (\ref{eq:u_2 example}).

%% file: Conclusion.tex
\section{Conclusion}\label{sec:conclusion}
This paper addresses the problem of making a class of nonlinear systems with  relative degree less than or equal to two nonlinear NI or OSNI using state feedback control. Roughly speaking, the system of interest is state feedback equivalent to a nonlinear NI or OSNI system with a positive definite storage function if and only if the system in question is weakly minimum phase. This result also helps achieving stabilization when the system in question has nonlinear ONSI uncertainty. The process of nonlinear NI state feedback equivalence and stabilization is illustrated by a numerical example.